\documentclass{article}   	
\usepackage{arxiv}
\usepackage{amssymb}
\usepackage{amsmath}

\usepackage{algpseudocode}
\usepackage{algorithm}
\usepackage{booktabs}
\usepackage{pifont}
\usepackage{multirow}

\usepackage{hyperref}
\hypersetup{
    colorlinks=true,
    linkcolor=blue,
    filecolor=magenta,
    citecolor=blue,
    urlcolor=blue,
    }
\renewcommand{\headeright}{}
\renewcommand{\undertitle}{}

\usepackage{sectsty}
\chapterfont{\fontfamily{qpl}\selectfont}
\sectionfont{\fontfamily{qpl}\selectfont}

\usepackage{authblk}

\usepackage[backend=biber,style=numeric-comp,sorting=none]{biblatex}
\usepackage{caption}
\usepackage{subcaption}
\usepackage{graphicx}
\usepackage[title]{appendix}
\usepackage{setspace}

\title{Woosh: A Sound Effects Foundation Model}
\author{
Gaëtan Hadjeres$^1$,\ \ Marc Ferras$^1$,\ \  Khaled Koutini$^1$,\ \ Benno Weck$^1$\thanks{Work done during an internship at Sony AI.},\ \ Alexandre Bittar$^{1*}$,\\
Thomas Hummel$^{1*}$,\ \ Zineb Lahrichi$^1$,\ \ Hakim Missoum$^1$,\ \ Joan Serrà$^1$,\ \ and Yuki Mitsufuji$^{1,2}$
}
\affil{\textmd{
$^1$ Sony AI\\ $^2$ Sony Group Corporation
}
}
\date{}							

\begin{document}
\fontfamily{qpl}\selectfont
\maketitle
\setcounter{footnote}{0}
\vspace*{-1.5cm}

\section*{Abstract}
The audio research community depends on open generative models as foundational tools for building novel approaches and establishing baselines. In this report, we present Woosh, Sony AI's publicly released sound effect foundation model, detailing its architecture, training process, and an evaluation against other popular open models. Being optimized for sound effects, we provide (1)~a high-quality audio encoder/decoder model and (2)~a text-audio alignment model for conditioning, together with (3)~text-to-audio
and (4)~video-to-audio generative models. Distilled text-to-audio and video-to-audio models are also included in the release, allowing for low-resource operation and fast inference. Our evaluation on both public and private data shows competitive or better performance for each module when compared to existing open alternatives like StableAudio-Open and TangoFlux. Inference code and model weights are available at \url{https://github.com/SonyResearch/Woosh}. Demo samples can be found at \url{https://sonyresearch.github.io/Woosh/}.

\tableofcontents    

\section{Introduction}

Current research on generative audio modeling largely focuses on conditioned models, mainly based on textual descriptions of audio. While significant progress has been made in terms of modeling, most approaches do not provide open weights for the research community to build upon~\cite{agostinelli23,evans24,evans24long,hung25}. Other approaches like~\cite{kreuk23,liu23} do provide open weights, but use low audio sampling rates only up to 16\,kHz. Notable exceptions are AudioLDM2~\cite{liu24}, StableAudio-Open~\cite{evans24sao}, and TangoFlux~\cite{hung25}, which generate higher quality audio while addressing the generation of both general audio and music. MusicGen~\cite{copet24} is another high-quality open model which specializes on music generation.

In this report, we propose a text-conditioned generative model specializing on instantaneous high-quality sound effect generation. Based on the multimodal FLUX-Kontext extension~\cite{flux25}, our latent diffusion model (LDM) has been optimized from the ground up for sound effects and for professional use. As part of the public release (\url{https://github.com/SonyResearch/Woosh/}), we provide inference code and open weights for non-commercial use for the full pipeline: encoder/decoder, text-conditioning, and diffusion models, the latter being further distilled for instant generation. We benchmark our public and private models against StableAudio-Open and TangoFlux, and our CLAP model against LAION-CLAP. For the best audio generation quality, a version trained exclusively on a large amount of studio-quality licensed sound effect libraries is internally available\footnote{Please contact \href{mailto:marc.ferrasfont@sony.com}{Marc Ferras} or \href{mailto:hakim.missoum@sony.com}{Hakim Missoum} for more information about this model.}.

\begin{figure}[t]
\centering
\begin{subfigure}{0.5\textwidth}
	\centering
	\includegraphics[scale=0.2]{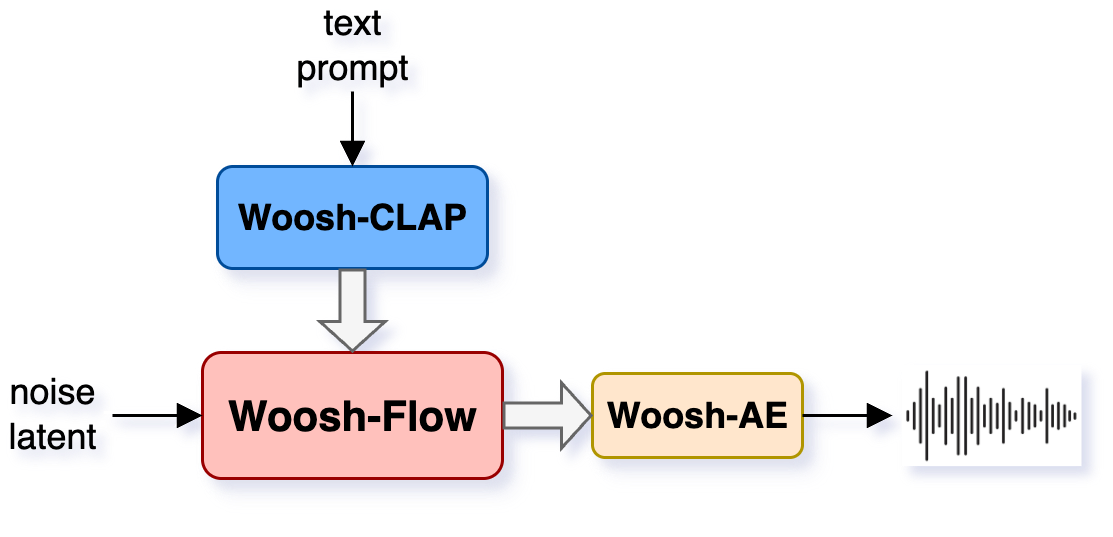}
\end{subfigure}%
\begin{subfigure}{0.5\textwidth}
	\centering
	\includegraphics[scale=0.2]{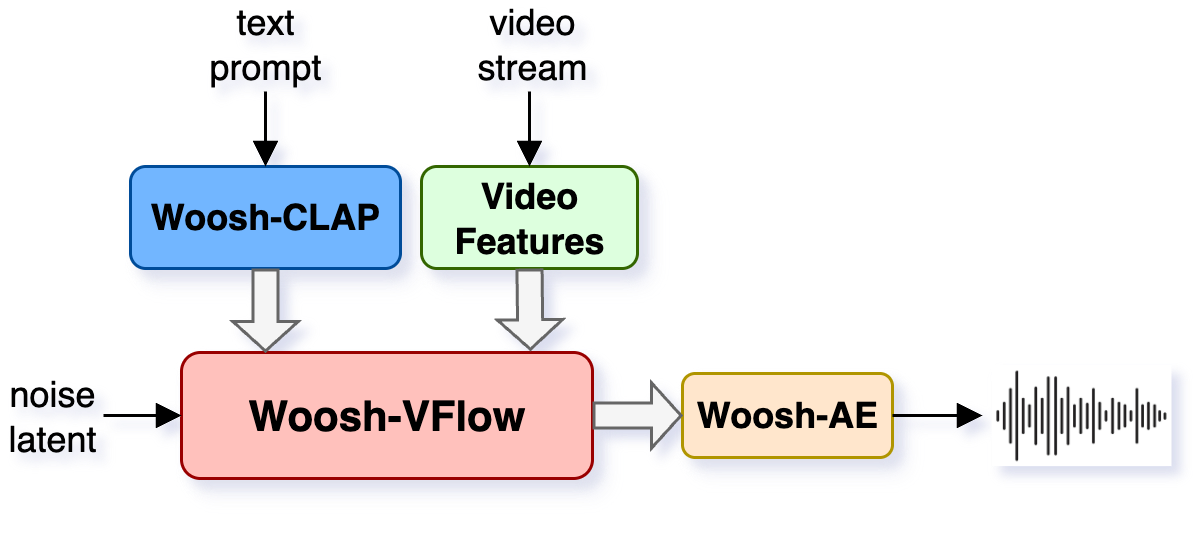}
\end{subfigure}
\caption{Inference-time layout of the Woosh-Flow (left) and Woosh-VFlow (right) models for text-to-audio and video-to-audio generation, respectively.}
\label{sfxfm}
\end{figure}

The current public release provides four models that address the text-to-audio~(T2A) and video-to-audio~(V2A) tasks:
\begin{itemize}
\item{\textbf{Audio encoder/decoder (Woosh-AE)}} --- High-quality latent encoder/decoder providing latents for generative modeling and decoding audio from generated latents.
\item{\textbf{Text conditioning (Woosh-CLAP)}} --- Multimodal text-audio alignment model providing token latents for diffusion model conditioning or CLAP scoring.
\item{\textbf{T2A Generation (Woosh-Flow and Woosh-DFlow)}} --- Original and distilled LDMs generating audio unconditionally or from a given a text prompt.
\item{\textbf{V2A Generation (Woosh-VFlow and Woosh-DVFlow)}} --- Multimodal LDM generating audio from a video sequence with optional text prompts.
\end{itemize}
Figure~\ref{sfxfm} shows a high-level diagram of the Woosh-Flow and Woosh-VFlow models performing T2A and V2A generation, respectively. Demo samples for the encoder/decoder, T2A, and V2A modules can be found at \url{https://sonyresearch.github.io/Woosh}.


\section{Woosh-AE: Audio Encoder/Decoder}
\label{sfxfm_encdec}

\subsection{Architecture}
The Woosh-AE module is based on the VOCOS architecture~\cite{siuzdak24}, a GAN-based vocoder operating on the domain of the short-time Fourier transform (STFT) complex coefficients. Compared to other popular encoder/decoders like Encodec~\cite{defossez22} or DAC~\cite{kumar23}, the VOCOS vocoder does not use quantization and it performs one-step down/up sampling via STFT/iSTFT efficiently. The iSTFT avoids aliasing artifacts, typically linked to transposed convolutions used in upsampling~\cite{pons21icassp} in alternative approaches. Note that Woosh-AE works on monaural audio only.

\begin{figure}[t]
\centering
\includegraphics[scale=0.2]{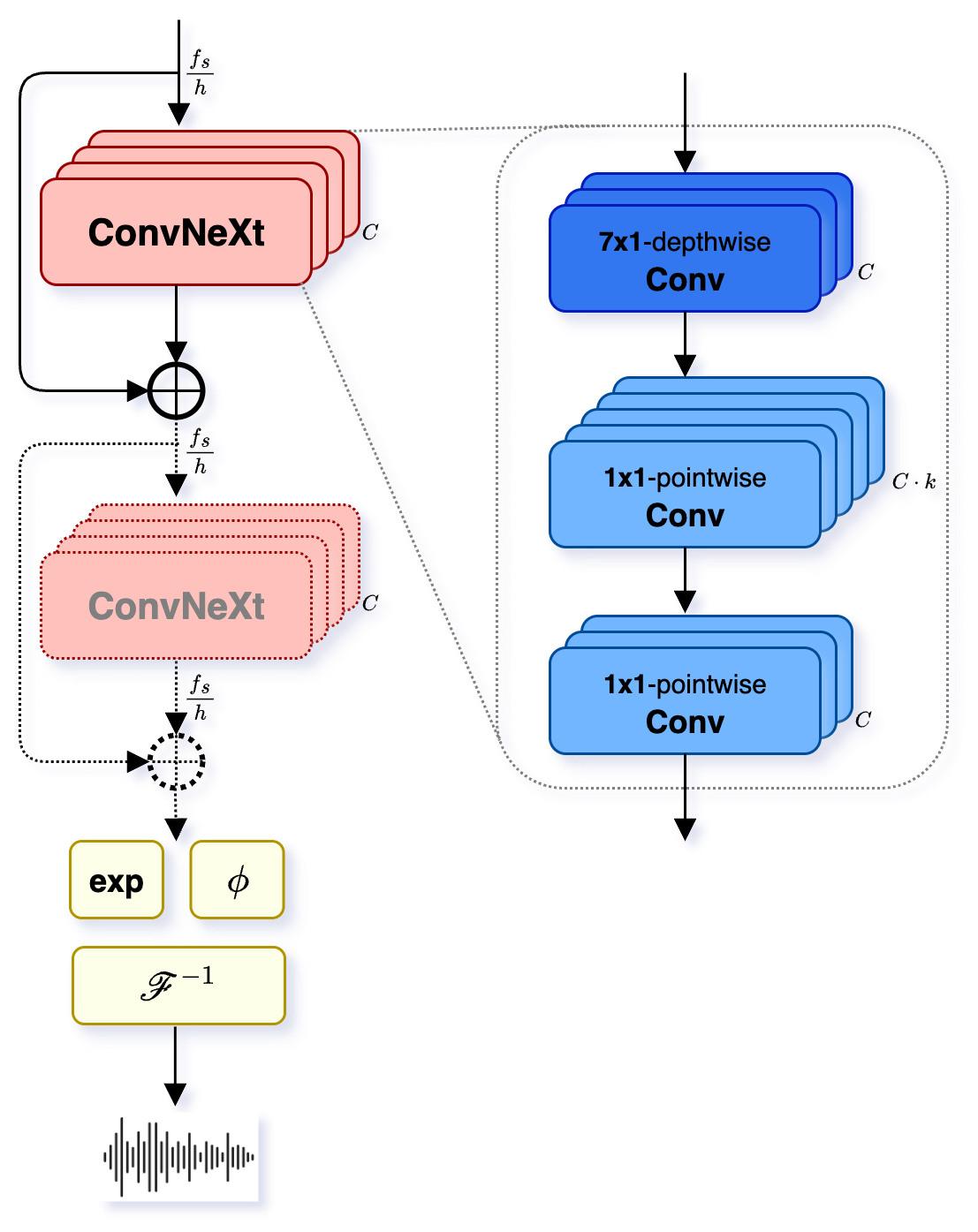}
\caption{VOCOS decoder architecture as a cascade on ConvNeXt blocks, used in Woosh-AE.}
\label{vocos-convnext}
\end{figure}

The VOCOS architecture uses a cascade of ConvNeXt blocks with residual connections both in the encoder and decoder~\cite{siuzdak24}, as shown in Figure~\ref{vocos-convnext}. The last ConvNeXt block is meant to predict magnitudes $\mathbf{m}$ and phases $\mathbf{p}$, from which the complex STFT is computed as $\mathbf{m}\cdot (\cos(\mathbf{p}) + j \sin(\mathbf{p}))$ = $\mathbf{m}\cdot(\mathbf{x} + j \mathbf{y})$, prior to STFT inversion. This formulation avoids phase wrapping issues for any phase vector $\mathbf{p}$. However, our model uses a slightly different approach to computing complex STFT coefficients. The last ConvNeXt block predicts magnitude $\mathbf{m}$, and real $\mathbf{x'}$ and imaginary $\mathbf{y'}$ parts instead of phase. Then $\mathbf{x'}$ and $\mathbf{y'}$ are used to compute the phase-only components on the complex plane as $\mathbf{x}=\mathbf{x'} / \mathbf{m'}$ and $\mathbf{y}=\mathbf{y'}/\mathbf{m'}$ components, with $\mathbf{m'}=\sqrt{ \mathbf{x'}^2 + \mathbf{y'}^2}$. The STFT coefficients are finally obtained as $\textrm{softplus}(\mathbf{m})\cdot(\mathbf{x} + j \mathbf{y})$. This is an alternative way of avoiding phase wrapping issues that improved reconstruction and discrimination metrics in preliminary experiments.

\subsection{Data}
\label{sec:ae_data}
To train the Woosh-AE public model, we used the following datasets:
\begin{itemize}
\item \textbf{Freesound}~\cite{font13} - A dataset consisting of Creative Commons licensed sound clips collected on the \url{freesound.org} online platform. Only audio with public domain (CC0), CC-BY, and CC-Sampling+ was used. The dataset contains a total of 370\,k audio files, with the vast majority sampled at least at 44.1\,kHz. We kept the original sample descriptions from Freesound.

\item \textbf{AudioCaps} --- A dataset with 48\,k audio clips from AudioSet~\cite{gemmeke17}, captioned by humans in a crowd-sourced setting~\cite{audiocaps}. The sample rate for all samples is 32\,kHz.

\item \textbf{WavCaps}~\cite{mei24} --- A dataset consisting of 99\,k weakly-labeled captions along with the corresponding Audioset audio clips only (WavCaps-AudioSet subset). Audio samples overlapping with AudioCaps were dropped. The sample rate is 32\,kHz.

\item \textbf{VTCK}~\cite{yamagishi2019vctk} --- A dataset consisting of 44\,k samples of accented English speech. Only the train split of the dataset was used, and only the audio was used, dropping any metadata or transcripts. Originally thought for high-quality speech synthesis and voice cloning, the sampling rate is 48\,kHz.

\item \textbf{Wapy} --- An internal synthetic dataset generated using phase-modulation synthesis. For each of the 100\,k audio samples, we generate a mix with 1 to 3 random audio signals, each with a random number of up to 20 sinusoidal partials, random interpolated pitch trajectories, random spectral envelopes, and random harmonic deviation. The main purpose behind this unlabeled dataset was to excite the entire range of frequencies during model training, with the hope of reducing audio distortion in the generated audio. The sample rate is 48\,kHz.

\item \textbf{Internal music (IM)} --- An internal dataset consisting of 78\,k commercially-licensed popular music songs. Single stems as well as mixed stems were used for training, with stem labels used only for the purpose of mixing. The sample rate is 44.1\,kHz.

\end{itemize}


To train the Woosh-AE-Private model, we used VCTK, Wapy, Internal Music and a mix of several commercially-licensed studio-quality sound effect libraries, involving around 1\,M samples and 5500\,h of commercial audio. The pre-processing for training both public and private autoencoders consisted of resampling the audio to 48\,kHz and randomly taking 1-second long chunks from each audio. Since the architecture is fully-convolutional, the model can operate with any length at inference time.

\subsection{Training}
Closely following DAC \cite{kumar23}, our model uses a GAN approach to training, minimizing a linear combination of the losses below:
\begin{itemize}
\item \textbf{Frequency-domain spectral loss} --- To encourage modeling at multiple time scales, multi-scale L1 losses are computed on mel spectrograms with window sizes $W=\{32, 64, 128, 256, 512, 1024, 2048\}$ and mel bin sizes $\{[5, 10, 20, 40, 80, 160, 320]\}$, as in~\cite{kumar23}. Assuming $G$ is the encoder-decoder model, and thus the generator in the adversarial training framework:
\begin{equation*}
\mathcal{L}_{spec} = \sum_{w \in W} \sum_t \left\| \mathcal{M}(\mathcal{S}_t(x)) - \mathcal{M}(\mathcal{S}_t(G(x))) \right\|_1 ,
\end{equation*}
where $\mathcal{M}$ is the frequency-to-mel conversion and $\mathcal{S}_t$ is the STFT at time $t$.

\item \textbf{Adversarial loss} --- A total of 8 discriminators are used for training: multi-period discriminators (MPD)~\cite{kong20} with periods $\{2,3,5,7,11\}$, and multi-band multi-scale complex STFT discriminators (MBMS-STFT)~\cite{kumar23,zeghidour21,defossez22} with window sizes $\{512,1024,2048\}$, each using 5~bands. The sum of discriminator outputs is used in a least squares GAN objective, as in~\cite{kumar23,mao17}, minimizing
\begin{equation*}
\begin{split}
\mathcal{L}_{LSGAN}(D) & = \frac{1}{2} \mathbb{E}_{x} \left[ (D(x) - 1)^2 \right] + \frac{1}{2} \mathbb{E}_x \left[ (D(G(x)) +1)^2 \right], \\
\mathcal{L}_{adv} & = \mathbb{E}_x \left[ (D(G(x)) - 1)^2 \right] .
\end{split}
\end{equation*}

\item \textbf{Feature matching loss} --- The L1 norm of the difference between discriminator feature maps for real and generated audio, taking expectation over all layers and discriminators, as
\begin{equation*}
\mathcal{L}_{FM}(G, D_k) = \mathbb{E}_{x} \left[ \sum_{i=1}^{L} \frac{1}{N_i}  \left\|  D_k^{(i)}(x) - D_k^{(i)} (G(x))  \right\|_1 \right] ,
\end{equation*}
where $G$ and $D_k^{(i)}$ are the generator and the $i$-th layer of the $k$-th discriminator, $N_i$ is the number of units in each layer, and $x$ and $G(x)$ are true and generated audio samples, respectively. Feature matching is used at each intermediate layer of each discriminator.

\end{itemize}

The final loss is computed as the weighted average
\begin{equation*}
\mathcal{L} = 15 \mathcal{L}_{spec} + \mathcal{L}_{adv} + 2 \mathcal{L}_{FM} .
\end{equation*}
The model uses 8 encoder/decoder layers with latent, main, and intermediate dimensions $d_{l}=128$, $d_{m}=2048$, and $d_{i}=3072$, respectively. The STFT analysis/synthesis uses $n_{\text{FFT}}=960$ and $n_{\text{hop}}=480$ as FFT and hop sizes, respectively. With audio being sampled at 48\,kHz, this results in an audio-to-latent compression ratio of 3.75.

\subsection{Evaluation}
\label{sec:ae_evaluation}
We evaluate the Woosh-AE model on the following public and private datasets:
\begin{itemize}

\item \textbf{AudioCaps ---} We use the test split of AudioCaps as public dataset for evaluation. It consists of 950 audio samples from AudioSet with crowd-sourced manual captions.

\item \textbf{InternalSFX ---} A subset of 5000 audio samples from the A Sound Effect studio-quality audio library not used for training. This dataset has detailed but rather non-verbose descriptions that are used as captions. As opposed to AudioCaps, most of the content consists of original sounds for production, rather than composed audio scenes recorded in real environments. We downsample the audio to 48\,kHz for consistency with the other datasets and for computational efficiency. This is referred to as our private evaluation data.

\end{itemize}

We compare Woosh-AE-Public and Woosh-AE-Private against the StableAudio-Open variational autoencoder (SAO-VAE) \cite{evans24sao}, and the Descript (16\,kbps) \cite{kumar23} and Encodec (24\,kbps) \cite{defossez22} baselines, the latter two being neural audio codecs. We compute the following metrics:
\begin{itemize}

\item \textbf{Log-mel Distance (MelDist) ---} The L1 distance between ground-truth audio and reconstructed audio log-mel spectra. 128 mel bands were used.

\item \textbf{Log-STFT Distance (STFTDist) ---} The L1 distance between ground-truth audio and reconstructed audio STFT magnitude spectra. An FFT size of 512 samples was used.

\item \textbf{Scale Independent Signal-to-Distortion Ratio (SI-SDR) ---} Ground truth audio norm to reconstruction error norm ratio, scaling the ground truth to match the reconstructed audio.

\end{itemize}

Table \ref{tab:ae_results} shows results for the encoder/decoder evaluation. For both AudioCaps and InternalSFX datasets, Woosh-AE-Public and Woosh-AE-Private ourperform all other models by a large margin. For AudioCaps, Woosh-AE-Public obtains reconstruction metrics 85\% lower in terms of MelDist (0.032 vs.\ 0.217) and 23\% lower in STFTDist (1.18 vs.\ 1.55) compared to SAO-VAE. The numbers also show a large drop in SI-SDR for SAO-VAE, although a negative SI-SDR is also reported for this dataset in \cite{evans24sao}. We suspect SI-SDR may not be a reliable metric for audio generated with a stochastic model such as SAO-VAE. The Descript neural audio codec, operating at 16\,kbps, obtains better metrics than SAO-VAE, which is not quantized, with an SI-SDR of 9.69\,dB, but still far from the 20.79\,dB for Woosh-AE-Public or the 22.52\,dB for Woosh-AE-Private. Encodec, the only baseline working at 48kHz sample rate, performs poorly overall. 

For the InternalSFX test set, both Woosh-AE models still perform comparably close, even though no commercial sound data was used for training Woosh-AE-Public. Absolute SI-SDR for these models are still high at 12.7\,dB and 16.03\,dB, but lower than for AudioCaps, suggesting that InternalSFX a is more demanding dataset (e.g.,~in terms of transients and high-frequency content). Compared to Woosh-AE-Private, SAO-VAE results in 82\% and 41\% lower MelDist (0.021 vs.\ 0.121) and STFTDist (0.272 vs.\ 0.462), respectively.

The metrics above, although favorable to Woosh-AE, need to be put into perspective. SAO-VAE uses a latent dimension of 64 at a frame rate of 21.5\,Hz, and the sample rate is 44.1\,kHz. The Woosh models use a latent dimension of 128 at an $\approx$5$\times$ faster frame rate of 100\,Hz, with audio being sampled at 48\,kHz. Thus, the compression ratio for SAO-VAE is 0.03 whereas for Woosh-AE it is 0.26 (i.e.,~9~times lower). Woosh-AE is also the largest model with 221\,M parameters, followed by SAO-VAE with 156\,M, and far above Descript or Encodec.

\begin{table}[t]
\centering
\caption{Woosh-AE evaluation on the public AudioCaps-Test and the InternalSFX-Test datasets.}
\vspace*{2mm}
\resizebox{\columnwidth}{!}{%
\begin{tabular}{lcccccccc}
\toprule
 &  & & \multicolumn{3}{c}{\textbf{AudioCaps-Test}} & \multicolumn{3}{c}{\textbf{InternalSFX-Test}}\\
\cmidrule{4-6} \cmidrule{7-9}
\textbf{Model} & \textbf{SampleRate} &  \textbf{Params} & \textbf{MelDist} &  \textbf{STFTDist} & \textbf{SI-SDR} & \textbf{MelDist} & \textbf{STFTDist} & \textbf{SI-SDR}\\
\toprule
Descript           & 44.1\,kHz & 76\,M  & 0.150 & 1.44 &  9.69 & 0.081 & 0.376 &  7.29 \\
Encodec            & 48\,kHz & 19\,M    & 0.266 & 2.00 &  7.79 & 0.161 & 0.784 &  5.22 \\
SAO-VAE            & 44.1\,kHz & 156\,M & 0.217 & 1.55 & $-$0.08 & 0.121 & 0.462 & $-$2.92 \\
\midrule
Woosh-AE-Public    & 48\,kHz & 221\,M   & \textbf{0.032} & \textbf{1.18} & 20.79 & \textbf{0.021} & \textbf{0.272} & 12.70 \\
Woosh-AE-Private   & 48\,kHz & 221\,M   & 0.074 & 1.25 & \textbf{22.52} & 0.024 & 0.288 & \textbf{16.03} \\

\toprule
\end{tabular}
}
\label{tab:ae_results}
\end{table}

\section{Woosh-CLAP: Text Conditioning}
\label{sfxfm_clap}
\subsection{Architecture}
The Woosh-CLAP text conditioning module uses a text encoder trained for text-audio alignment following the contrastive language-audio pretraining (CLAP)~\cite{elizalde22} approach. CLAP uses text and audio encoder models that produce text and audio embeddings, respectively. During training with a multi-modal dataset, positive text-audio embedding pairs are pushed together and negative pairs are pulled apart. As a result, text and audio embeddings end up living on the same vector space, so-called \textit{shared space}. This space, where audio can be predicted from text and vice-versa, turns out to be highly semantic.

Woosh-CLAP uses RoBERTa-Large~\cite{liu19} as the text encoder and PaSST~\cite{koutini22} as the audio encoder. A linear head is used to project audio- and text-pooled embeddings to a shared space of dimension 1024. RoBERTa has 355\,M parameters, 24~transformer layers, 16 attention heads, and a hidden dimension of 1024. PaSST uses patched spectrogram inputs and a similar architecture to vision transformers ~\cite{dosovitskiy21}, with 86\,M parameters, 12~transformer blocks, 12 attention heads, and a hidden dimension of 768. PaSST has some desirable properties, such as being able to handle long audios by aggregating per-chunk CLAP embeddings. 
The audio encoder is dropped after training, as it is not required for text embedding computation and generation. However, in our release we provide both text and audio encoders to promote the reuse of Woosh-CLAP in further applications even beyond generation.

\subsection{Data}
To train the Woosh-CLAP-Public model, we used the Freesound, AudioCaps, and WavCaps datasets described in Section~\ref{sec:ae_data}. The Woosh-CLAP-Private model was trained using only the same mix of commercial studio-quality sound effect libraries as described in the same section. Compared to the public datasets, the captions provided by these internal datasets were clearly less verbose, keyword-like, and using specific technical vocabulary and abbreviations. We used a large language model (LLM) to generate alternative descriptions that were more verbose and closer to natural language.

Pre-processing consisted of resampling the audio down to 32\,kHz (the PaSST operating sampling rate), and training was done by taking random, non-segmented 10-second long chunks from the full audio sample.

\subsection{Training}
The official RoBERTa-Large and PaSST models, pretrained on multiple tasks using large amounts of heterogeneous text and audio data, served as a solid fine-tuning base. Linear text and audio heads were appended to the encoders as a projection from their respective embedding dimensions to the shared space dimension. Fine-tuning helps reduce text and audio domain shifts found in sound effect audio and in captions in particular.

\begin{figure}[t]
\centering
\includegraphics[scale=0.3]{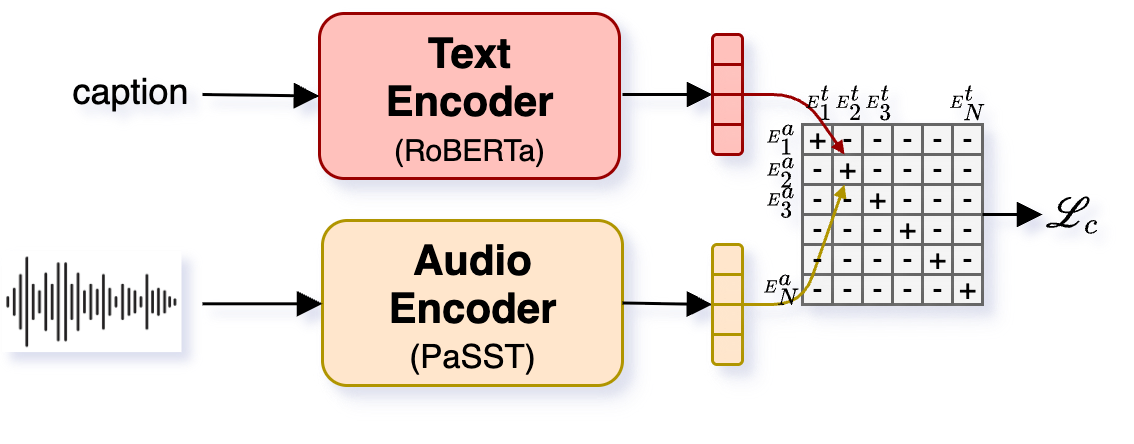}
\caption{Woosh-CLAP training block diagram for a positive pair of samples. Only the text encoder is used at generation time as conditioning. }
\label{clap-training}
\end{figure}

Assuming the $i$-th training sample pair having a text caption $\mathbf{X}^t_i$ and an audio sample $\mathbf{X}^a_i$, we project to the shared space by
\begin{eqnarray*}
 E_i^t = \mathbf{W}^t f(\mathbf{X}^t_i) + \mathbf{b}_t ,\\
 E_i^a = \mathbf{W}^a f(\mathbf{X}^a_i) + \mathbf{b}_a .
\end{eqnarray*}
The symmetric contrastive loss~\cite{radford21} is then computed as
\begin{equation*}
 L = \frac{1}{2N} \sum_{i=1}^{N} \left( \log \frac{\exp\left(E_{i}^{a} \cdot E_{i}^{t} / \tau\right)}{\sum_{j=1}^{N} \exp\left(E_{i}^{a} \cdot E_{j}^{t} / \tau\right)} + \log \frac{\exp\left(E_{i}^{t} \cdot E_{i}^{a} / \tau\right)}{\sum_{j=1}^{N} \exp\left(E_{i}^{t} \cdot E_{j}^{a} / \tau\right)} \right) ,
\end{equation*}
where $N$ is the batch size and $\tau$ is a temperature value. The numerators are computed for the positive pairs, i.e.,~the diagonal elements in the cosine similarity matrix of Figure~\ref{clap-training}, and the denominators are computed for all pairs, i.e.,~all cosine similarity elements.

We compute the contrastive loss on gathered text and audio embeddings over 4 GPUs, each with batch size 112, resulting in an effective batch size of $N=448$ samples. Increasing the effective batch size did not have a positive effect on retrieval metrics. A fixed temperature of $\tau=0.2$ was used.

\subsection{Evaluation}

We conduct the evaluation on the same public and private datasets used in Section~\ref{sec:ae_evaluation} as they highlight the influence of the data choice on the reported metrics. We evaluated three CLAP models: 1) LAION-CLAP~\cite{wu24}, using RoBERTa+HTS-AT encoders~\cite{liu19,chen22}, 2) Woosh-CLAP trained on public datasets, and 3) Woosh-CLAP trained on private studio-quality datasets. We assessed their performance in terms of recall in text-to-audio (T2A) and audio-to-text (A2T) retrieval tasks, and on the aforementioned AudioCaps and InternalSFX datasets.

Table~\ref{tab:clap} shows different recall metrics for the three aforementioned models. Considering the public test set, LAION-CLAP outperforms Woosh-CLAP-Public by up to 13\% in T2A-R@10 (0.546 vs.\ 0.618).  Considering that both models use similar architectures and training criterion, and that LAION-CLAP uses a non-negligible proportion of commercial training data, we attribute the observed performance gap to the training dataset choices and the amount of training data. Woosh-CLAP-Private, trained exclusively on professional audio libraries compares poorly to both LAION-CLAP and Woosh-CLAP-Public models. However, when considering the internal dataset evaluation consisting of studio-quality sound effect data, we see a totally different picture (recall that InternalSFX-Test is independent of the training sets). In this setting, results for LAION-CLAP and Woosh-CLAP-Public models lie far behind Woosh-CLAP-Private, the latter obtaining a relative improvement up to 248\% higher over LAION-CLAP in T2A-R@10 (0.655 vs.\ 0.188). A similar trend is observed for A2T retrieval, where LAION-CLAP and Woosh-CLAP-Public also perform poorly, with Woosh-CLAP-Private obtaining over 2~times higher A2T-R@10 (0.766 vs.\ 0.328).

These results highlight the importance of using commercial audio libraries when targeting professional applications. In our experience, while in-the-wild audio can sometimes be of recording quality that is comparable to open datasets, its content can be structured very differently.   On the other hand, public audio data tend to be real recordings, involving a mix of source sounds, making it difficult to disentangle each sound for production purposes. They are often also unintentionally noisy. The annotation style provided in professional libraries can also be highly mismatched to those provided in public datasets. We believe these factors explain the disparity observed in the reported evaluation results to a large extent.

\begin{table}[t]
\centering
\caption{Woosh-CLAP evaluation on the public AudioCaps-Test and the InternalSFX-Test datasets. Recall at 5 (R@5) and recall at 10 (R@10) are given for text-to-audio (T2A) and audio-to-text (A2T) retrieval.}
\vspace*{2mm}
\begin{tabular}{lccccccccc}
\toprule
\textbf{Model} & \textbf{Params} &  \multicolumn{4}{c}{\textbf{AudioCaps-Test}} & \multicolumn{4}{c}{\textbf{InternalSFX-Test}}\\
\cmidrule{3-6} \cmidrule{7-10}
& & \multicolumn{2}{c}{T2A} & \multicolumn{2}{c}{A2T} & \multicolumn{2}{c}{T2A} & \multicolumn{2}{c}{A2T}\\
 &  &  R@5 &  R@10 & R@5 &  R@10 & R@5 &  R@10 & R@5 & R@10\\
\toprule
LAION-CLAP         & 355\,M + 31\,M & \textbf{0.469} & \textbf{0.618} & \textbf{0.467} & \textbf{0.650} & 0.280 & 0.188 & 0.215 & 0.328\\
Woosh-CLAP-Public  & 355\,M + 86\,M & 0.428 & 0.546 & 0.462 & 0.595 & 0.138 & 0.198 & 0.281 & 0.401\\
Woosh-CLAP-Private & 355\,M + 86\,M & 0.095 & 0.167 & 0.153 & 0.227 & \textbf{0.592} & \textbf{0.655} & \textbf{0.694} & \textbf{0.766} \\
\toprule
\end{tabular}
\label{tab:clap}
\end{table}

\section{Woosh-Flow: Text-to-Audio Generation}
\label{sfxfm_t2a}

\subsection{Architecture}
\begin{figure}[t]
\centering
	\includegraphics[scale=0.2]{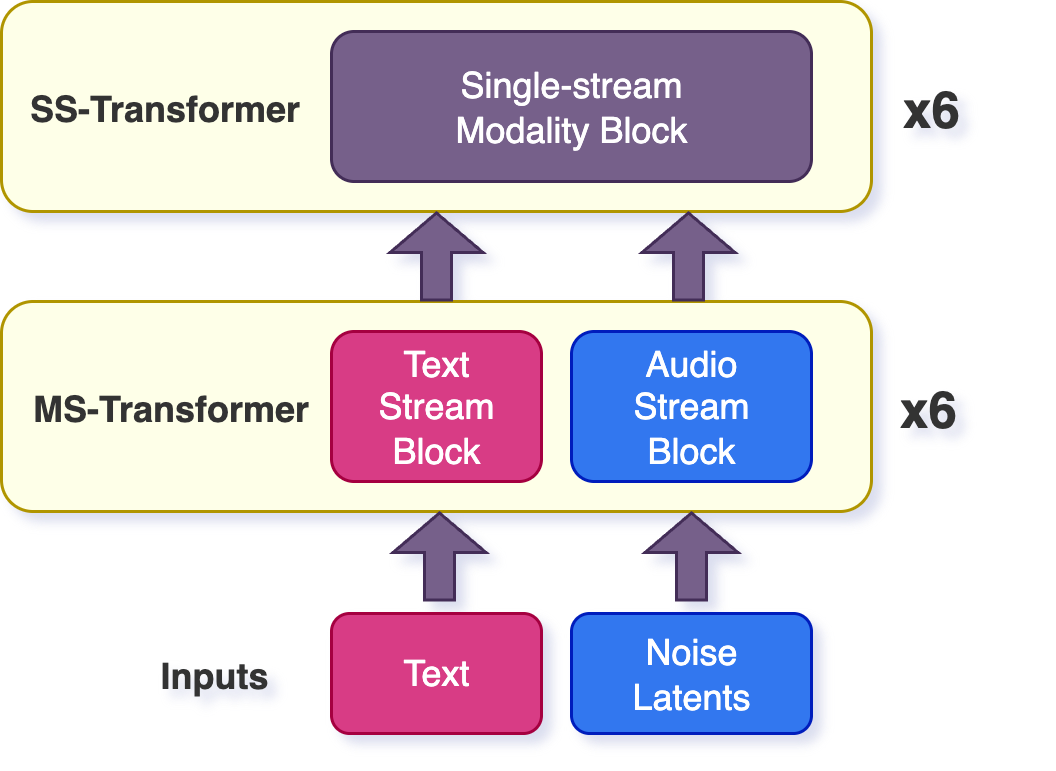}
	\caption{Multimodal transformer stack in the Woosh-Flow diffusion model, formed by MultiStream (MS) and SingleStream (SS) blocks.}
	\label{flux-arch}
\end{figure}

\begin{figure}[!h]
\centering
\begin{subfigure}{0.5\textwidth}
	\centering
	\includegraphics[scale=0.15]{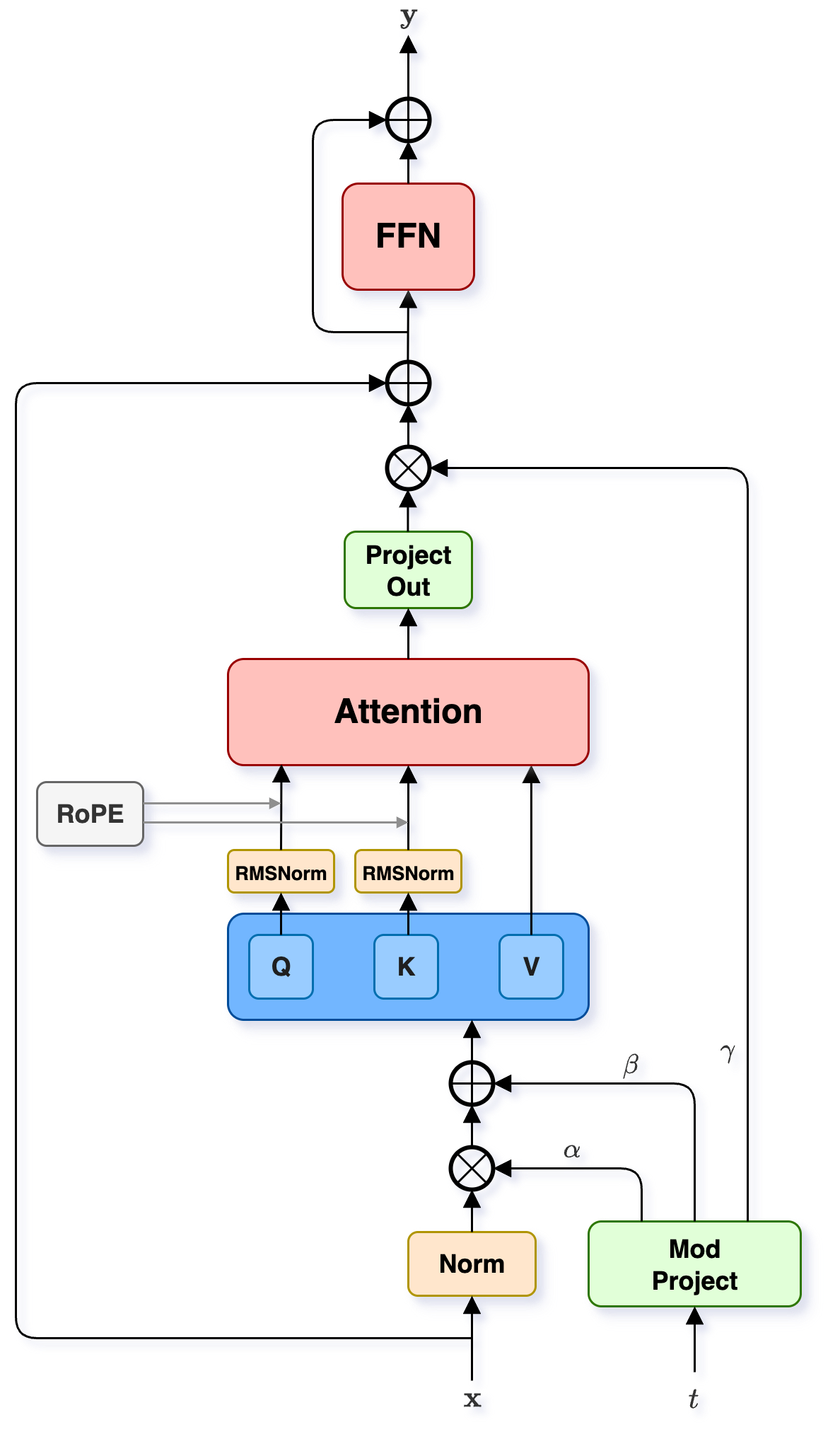}
\end{subfigure}%
\begin{subfigure}{0.5\textwidth}
	\centering
	\includegraphics[scale=0.15]{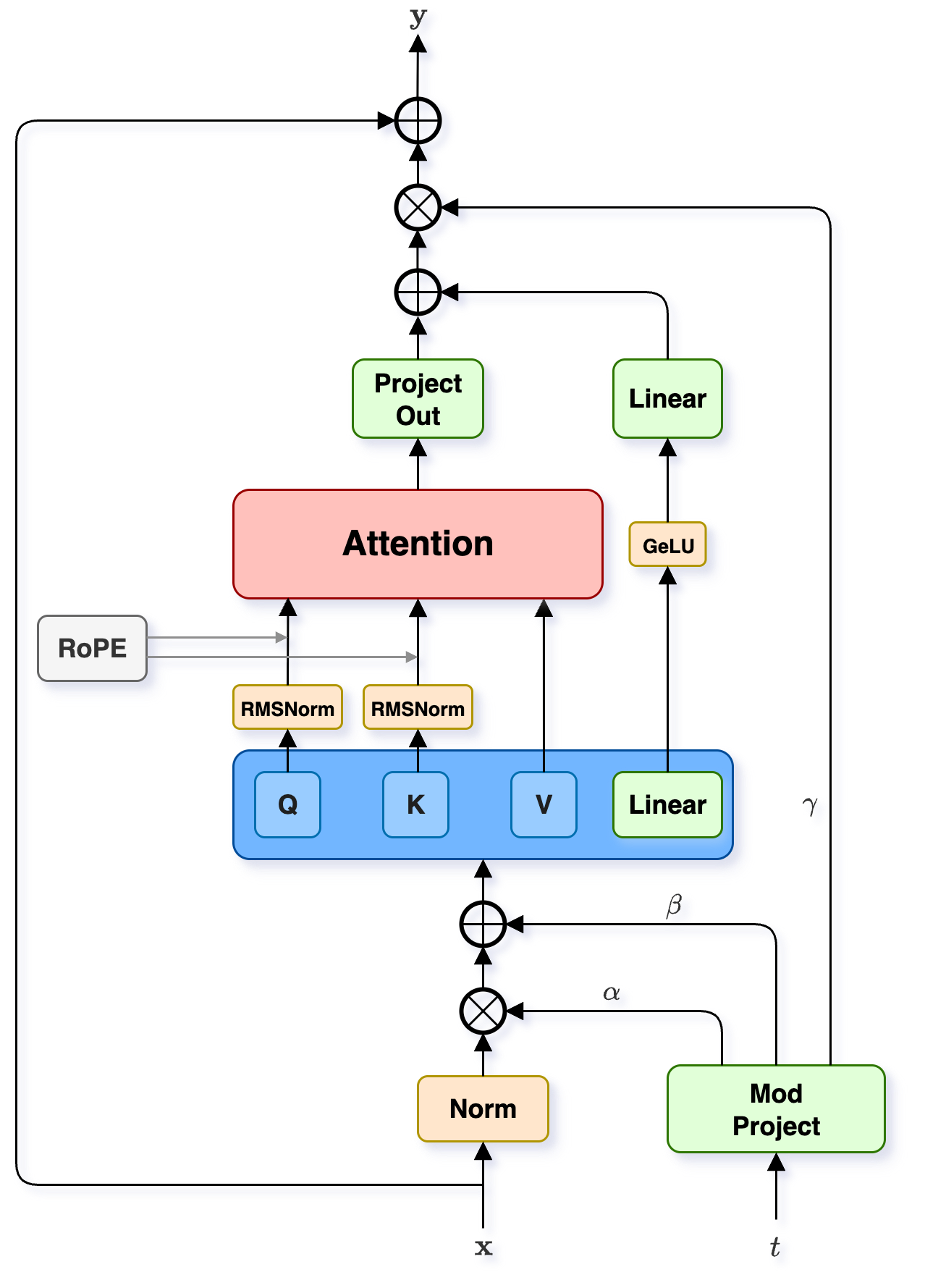}
\end{subfigure}
\caption{MultiStream transformer block diagram (left). Both self-attention and feed-forward network (FFN) outputs are computed independently for each modality sequence. SingleStream transformer block diagram (right). Self-attention is computed on the concatenation of modality sequences along the time dimension. Instead of an FFN block, non-linear features are in a parallel path and added after the attention project layer.}
\label{flux-mmmss}
\end{figure}

The Woosh-Flow generative model is a latent diffusion model operating on audio-encoded latents, extracted using the Woosh-AE model described in Section~\ref{sfxfm_encdec}. It is based on the FLUX-Kontext architecture~\cite{flux25}, using multimodal diffusion transformer blocks, and trained with the flow matching objective~\cite{lipman23}. The model uses a stack of multimodal transformer blocks, handling text and noise latent modalities, as shown in Figure~\ref{flux-arch}. MultiStream modality blocks compute both self-attention and feed-forward network outputs on each modality sequence independently, and use rotary positional embeddings (RoPE)~\cite{su23}. SingleStream modality blocks compute self-attention on a single sequence of time-concatenated modality sequences, also using RoPE. Figure~\ref{flux-mmmss} shows detailed diagrams for each of these blocks. SingleStream blocks attending to any time step across any modality sequence result in implicit attention computation across modalities, which can be thought of as indirectly conditioning the diffusion model.

The chosen architecture has a total of 12 multimodal transformer blocks, 6 MultiStream and 6 SingleStream. The main, intermediate, and RoPE dimensions are $d_m=1024$, $d_i=4096$, and $d_r=112$, respectively.

\subsection{Data}
To train the Woosh-Flow-Public model we use the Freesound, AudioCaps, and WavCaps datasets described in Section~\ref{sec:ae_data}. To train the Woosh-Flow-Private model we use the same mix of commercial studio-quality sound effect libraries described in the same section. Pre-processing consisted of resampling the audio to 48\,kHz. Audio samples were pre-segmented for non-silence using an energy-based detector, joining contiguous segments closer than 2 seconds. Random 5-second long chunks from a random segment were taken for training and testing.

\subsection{Training}
In a first stage, Woosh-Flow is trained using a flow matching (FM) objective~\cite{lipman23,lipman24}. The model is meant to estimate a velocity field $v_{t}$ that matches straight probability paths from data to noise. Assuming a linearly interpolated diffusion process of the form
\begin{equation*}
x_t = (1 - t)\cdot x_0 + t\cdot x_1 \, ,
\end{equation*}
with $x_0$ and $x_1$ being data and noise samples, respectively, its time derivative
\begin{equation}
\label{eq_dxdt}
v_t=\frac{d x_t}{dt} = x_1 - x_0
\end{equation}
is used as target velocity for the FM L2 loss:
\begin{equation}
    \mathcal{L}_{\textrm{FM}}(\theta) = \mathbb{E}_{x_0, x_1, t} \left\| u_\theta(x_t, t) - (x_1 - x_0) \right\|^2\, .
\label{eq_cfm_loss}
\end{equation}

The diffusion model network $u_\theta(x_t, t)$ is conditioned, as seen in the building blocks of Figure~\ref{flux-mmmss}, on both a noisy sample $x_t$ and its corresponding time step $t$, and approximates the target velocity $x_1 - x_0$. By learning this velocity field, the trained model is used as a denoiser to estimate the original sample $x_0$ from any noisy intermediate $x_t$ by solving the related differential equation $\frac{d x_t}{dt} = -u_\theta(x_t, t)$.

The criterion in Eq.~\ref{eq_cfm_loss} is used to pretrain the diffusion model for 300 epochs, using the Adam optimizer with 1000 warmup steps, $10^{-4}$ learning rate, and gradient clipping value at 1.0. The time step $t$ is sampled from a uniform distribution in the range $[0.001,1]$. The model parameters are smoothed with an exponential moving average and a 0.999 decay. For text-conditioned diffusion, the token sequence of the second last layer of the Woosh-CLAP text encoder (Section~\ref{sfxfm_clap}) is used as conditioning signal.

At inference time, the learned dynamics via velocity estimation define a well-specified initial-value problem, for which arbitrary ordinary differential equation (ODE) solvers can be used, e.g.,~higher-order adaptive solvers such as Dormand–Prince (DOPRI5). In practice, a trade-off between reconstruction quality and number of function evaluations (NFEs) is desired.

\subsection{Distillation}
\label{sec_t2a_distillation}
As a second training stage, the pretrained diffusion model is further distilled using the MeanFlow (MF) approach~\cite{geng25}, allowing to reduce the NFEs at inference time from 100 to 4 with negligible audio quality loss. We call this model Woosh-DFlow.

Instead of the FM instantaneous velocity target of Eq.~\ref{eq_dxdt}, MF introduces the average velocity $u_{t,r}$, defined as
\begin{equation}
\label{eq_ut1}
    u_{t,r}=\frac{1}{t-r}\int_{r}^t v_{\tau}d\tau \, ,
\end{equation}
that is, the average of the velocities in the interval $[r,t]$, and it is used as the velocity field target in Eq.~\ref{eq_cfm_loss}. Differentiating both sides of Eq.~\ref{eq_ut1} yields the identity
\begin{equation*}
    u(x_t,t,r)=v(x_t,t)-(t-r)\cdot\frac{d}{dt}\,u(x_t,t,r) \, ,
\end{equation*}
which, replaced as velocity target into the FM loss, yields the loss function
\begin{equation}
\mathcal{L}_{\textrm{MF}}(\theta) =  \mathbb{E}_{x_0, x_1, t, r} \left\| u_\theta(x_t, t, r) - (x_1 - x_0) + (t-r)\cdot\frac{d}{dt}\,u_{\theta^-}(x_t,t,r) \right\|^2 .
\label{eq_mf_loss}
\end{equation}
Here, $u_{\theta^-}$ in the last term indicates that gradients are stopped to avoid higher order derivatives. The Jacobian vector product (JVP) operation is used to evaluate $\frac{d}{dt}\,u_{\theta^-}(x_t,t,r) $.

In our experience, training the diffusion model from scratch using Eq.~\ref{eq_mf_loss} proved to be slow, with the JVP operation being computationally expensive, even when backing off to the FM criterion for a large proportion of training samples, as proposed in~\cite{geng25}. We switched to distillation mode, where the velocity target is the prediction by the teacher model $v_{\text{tgt}}(x_t,t)$ instead of $(x_1-x_0)$, yielding the distillation loss
\begin{equation}
\mathcal{L}_{\textrm{MF}}^{\text{d}}(\theta) =  \mathbb{E}_{x_0, x_1, t, r} \left\| u_\theta(x_t, t, r) - v_{\text{tgt}}(x_t,t) + (t-r)\cdot\frac{d}{dt}\,u_{\theta^-}(x_t,t,r) \right\|^2 .
\label{eq_mf_distill_loss}
\end{equation}

Considering that Woosh-Flow is a text-conditioned model, the MF procedure can also account for classifier-free guidance (CFG)~\cite{ho22} at training time. This enables the model to additionally save the unconditional evaluation required for CFG at inference time, and thus half of the NFEs. For this purpose, the MF distillation loss in Eq.~\ref{eq_mf_distill_loss} is further modified to evaluate the teacher target velocity with the CFG already applied, i.e.,~$v'_{\text{tgt}}(x_t,c,t) = (1-w) \cdot v_{\text{tgt}}(x_t,c,t) + w \cdot v_{\text{tgt}}(x_t,t)$, and taking the expectation over the distribution of CFG weights of interest as well:
\begin{equation*}
\mathcal{L}_{\textrm{MF}}^{\text{d,CFG}}(\theta) =  \mathbb{E}_{x_0, x_1, t, r,w} \left\| u_\theta(x_t, c, t, r) - v'_{\text{tgt}}(x_t,c,t) + (t-r)\cdot\frac{d}{dt}\,u_{\theta^-}(x_t,t,r) \right\|^2 .
\end{equation*}

To further improve the distilled model even more, we add an adversarial objective during distillation, using a discriminator alongside the generator. Based on the latent adversarial diffusion distillation approach~\cite{sauer24}, the discriminator learns to distinguish noisy ground truth (i.e.,~true samples), from noisy student-predicted (i.e.,~fake samples). The frozen teacher model is used as discriminator with the only addition of several convolutional heads that directly operate on the latent representations from different layers. The GAN hinge loss~\cite{lim17} is used in the discriminator together with adversarial losses. We adopt a similar approach to SANA~\cite{chen25} to compute partially denoised samples, except we directly denoise from $t$ to $r<t$ as $x_r^{\text{fake}}=x_t - (t-r)\cdot\text{student}(x_t, t, r)$, instead of denoising from $t$ to 0 and then re-noise to $r$ as done in SANA. We provide a detailed pseudocode for this training process in Algorithm~\ref{alg_ladd} in Appendix~\ref{appendix:ladd}.

We set the number of distillation fine-tuning epochs to 50, with a learning rate of $5\cdot10^{-6}$. To stabilize training in the first epochs, the MeanFlow joint $(t,r)$ embeddings were fine-tuned for 1000~steps to match the $t$ embeddings in the teacher network. CFG is uniformly sampled in the range [1,9] with a condition dropout rate of 0.1. The discriminator uses 4 convolutional heads. The adversarial loss weight is set to 0.5. For inference, a generic ODE solver, namely DOPRI5, is used with target absolute and relative tolerances of $10^{-3}$.

\subsection{Evaluation}
\label{sec:flow_evaluation}

We compare audio fidelity and semantic alignment for the Woosh-Flow and Woosh-DFlow models with SAO and TangoFlux, on the public and private datasets of Section~\ref{sec:ae_evaluation}, using the following metrics:
\begin{itemize}
\item \textbf{Audio quality and fidelity ---} To measure the perceptual quality of the generated audio, we compute the Fréchet distance between the OpenL3 embeddings of the generated and the ground truth audio samples (FD)~\cite{pascual23}. Additionally, we use the Kullback-Leibler divergence between the PaSST embeddings~\cite{koutini22} (or class probabilities) of the generated audio and the corresponding ground truth (KL), quantifying how well the model aligns acoustic and semantic content.

\item \textbf{Cross-modal alignment ---} We evaluate the alignment between the generated audio and the input conditioning modalities using the LAION-CLAP~\cite{WuCZHBD23laionclap} score (CLAP), assessing text-to-audio relevance by computing the cosine similarity between the target text caption and the generated audio. 
\end{itemize}

\begin{table}[t]
\centering
\caption{Woosh-Flow and Woosh-DFlow results on AudioCaps-Test and InternalSFX-Test. We provide the number latent diffusion model parameters (Params), the number of neural function evaluations at inference time (NFEs), and the FD, KL, and CLAP metrics described in Section~\ref{sec:flow_evaluation}. All models are conditioned on text.}
\vspace*{2mm}
\label{tab:results:flow_dflow}
\setlength{\tabcolsep}{9pt}
\resizebox{\columnwidth}{!}{%
\begin{tabular}{lccrrrrrrrr}
\toprule
\textbf{Approach} & \textbf{Params} & \textbf{NFEs} & \multicolumn{3}{c}{\textbf{AudioCaps-Test}} & \multicolumn{3}{c}{\textbf{InternalSFX-Test}} \\
\cmidrule{4-6}\cmidrule{7-9}
 &  &  & FD $\downarrow$ & KL $\downarrow$ & CLAP $\uparrow$ & FD $\downarrow$ & KL $\downarrow$ & CLAP $\uparrow$ \\
\midrule
SAO            & 1057\,M & ~~~2$\times$100 & 150.1 & 4.068 & 0.1457 & 341.0 & 2.690 & 0.2544 \\
TangoFlux      & 515\,M  & ~~~2$\times$100 & 131.9 & 1.700 & 0.3471 & 378.1 & 2.729 & 0.2249 \\
\midrule
Woosh-Flow-Public & ~~337\,M & $\approx$2$\times$70 & \textbf{109.1} & \textbf{1.599} & \textbf{0.3719} & 282.7 & \textbf{2.248} & \textbf{0.3752} \\
Woosh-DFlow-Public & ~~337\,M & 4 & 132.1 & 1.836 & 0.3229 & 322.5 & 2.404 & 0.3093\\
\midrule
Woosh-Flow-Private & ~~337\,M & $\approx$2$\times$70 & 213.8 & 4.238 & 0.1824 & \textbf{246.9} & 2.348 & 0.3010\\
Woosh-DFlow-Private & ~~337\,M & 4 & 276.2 & 4.748 & 0.1541 & 270.9 & 2.651 & 0.2569\\
\bottomrule
\end{tabular}
}
\end{table}

The inference setup for the considered models was set to be as close as possible. We use a CFG scale of 7, 100 sampling steps for SAO and TangoFlux, and a relative tolerance of $10^{-3}$ for Woosh-Flow, resulting in around 70 DOPRI5 solver steps. Note that, for SAO, TangoFlux, and Woosh-Flow, CFG forces two function evaluations, conditional and unconditional, per step, resulting in around 200, 200, and 140\,NFEs, respectively, to completing inference. For Woosh-DFlow, we use a CFG scale of 7 and a first-order Euler solver, with 4 sampling steps, also corresponding to the total number of NFEs for completing inference. The $t$-step scheduler is uniform and we use a renoising strategy, introducing noise at every sampling step, with weights $[0, 0.5, 0.5, 0.3]$.


Table~\ref{tab:results:flow_dflow} shows Frechet Distance (FD), KL-PaSST (KL) and LAION-CLAP score (CLAP) for the generated audio from Woosh models, together with SAO and TangoFlux, two other popular open-souce audio generative models. For AudioCaps, Woosh-Flow-Public outperforms TangoFlux and SAO, with 17\% and 27\% relative improvement in FD, and 6\% and 150\% relative improvement in CLAP score, respectively. Woosh-Flow-Private performs poorly on AudioCaps, but it still provides better text-audio alignment compared to SAO in terms of CLAP score, and performs comparably to TangoFlux. The Woosh-DFlow models follow closely their Woosh-Flow counterparts, while staying behind in all evaluation metrics. On the InternalSFX test set, Woosh-Flow-Private and Woosh-DFlow-Private outperform all other models, except for the Woosh-Public counterparts in some metric. In terms of FD, Woosh-Flow-Private is considerably better than all others; 27\% lower than SAO and 34\% lower than TangoFlux. Note that the KL and CLAP metrics internally use models trained on public audio data and captions, and we hypothesize that the ability to measure improvements for commercial data is inherently biased. For KL and CLAP score, the Woosh-Flow-Public model obtains the best scores, with 47\% (0.3752 vs.\ 0.2544) and 66\% (0.3754 vs.\ 0.2249) relative improvement in CLAP score over SAO and TangoFlux, respectively.

\section{Woosh-VFlow: Video-to-Audio Generation}
\label{sfxfm_v2a}

\subsection{Architecture}

Finally, we extend the pre-trained Woosh-Flow framework to support video-to-audio generation by introducing a video conditioner module and adapting the transformer architecture for a third modality. Specifically, the video conditioner utilizes features extracted from the input video using SynchFormer~\cite{iashin24} at a frame rate of 24\,Hz. These embeddings are projected via a linear layer to match the hidden dimension of the diffusion transformer, resulting in a sequence of video condition tokens.

To integrate these tokens, we modify the pre-trained multimodal transformer blocks to accommodate a dedicated video modality alongside the existing text and noise latent streams. We augment the MultiStream blocks by adding learnable projections for query, key, and value representations for each modality $m \in \{T, V, A\}$ (representing text, video, and audio/noise, respectively):
\begin{equation*}
\mathbf{Q}_m = \mathbf{X}_m \mathbf{W}^Q_m, \quad\quad \mathbf{K}_m = \mathbf{X}_m \mathbf{W}^K_m, \quad\quad \mathbf{V}_m = \mathbf{X}_m \mathbf{W}^V_m,
\end{equation*}
where $\mathbf{X}_m$ is the input sequence and $\mathbf{W}^Q_m, \mathbf{W}^K_m, \mathbf{W}^V_m \in \mathbb{R}^{d \times d}$ are modality-specific weight matrices. We perform joint self-attention on the concatenated sequences of all three modalities:
\begin{equation*}
[\mathbf{Z}_T \parallel \mathbf{Z}_V \parallel \mathbf{Z}_A] = \text{softmax}\left(\frac{\mathbf{Q}_{\text{joint}} \mathbf{K}_{\text{joint}}^\top}{\sqrt{d}}\right) \mathbf{V}_{\text{joint}} ,
\end{equation*}
where $\mathbf{Q}_{\text{joint}} = [\mathbf{Q}_T \parallel \mathbf{Q}_V \parallel \mathbf{Q}_A]$ (and similarly for $\mathbf{K}_{joint}$ and $\mathbf{V}_{joint}$), with $\parallel$ here denoting concatenation along the sequence dimension. Finally, we apply a distinct feed-forward network (FFN) with layer norm (LN) per modality by adding a new FFN specific to the video modality to process the attention outputs $\mathbf{Z}_m$:
\begin{equation*}
\mathbf{Y}_m = (\mathbf{X}_m + \mathbf{Z}_m) + \text{FFN}_m(\text{LN}(\mathbf{X}_m + \mathbf{Z}_m)) .
\end{equation*}

\subsection{Data}
\label{sec:data_vflow}
To train the video-to-audio generation model, we utilize two large-scale audio-visual datasets: VGGSound and OGameData250k (see below). A key challenge in training on these datasets is the alignment between the audio and the text descriptions. Therefore, we employ an audio-captioning pipeline using the Qwen3-Omni~\cite{qwen3omni} audio-language model to enrich the training supervision.
\begin{itemize}
\item \textbf{VGGSound ---}
The VGGSound \cite{Chen20vggsound} dataset contains 183\,k video clips for training, each of 10\,s in duration. While extensive, the dataset presents challenges regarding data quality: the visual content is often noisy or loosely related to the audio (e.g.,~containing unrelated background music or speech), and the original labels are frequently inaccurate, often describing only a fraction of the auditory scene or isolated acoustic events. To address this, we generated synthetic captions for the dataset using Qwen3-Omni. We processed the audio in 5-second windows with a 2.5-second hop size to capture its temporal evolution. During training, we employ a mixing strategy: we randomly select either the generated caption or the original dataset label with a probability of 0.5. Synthetic captions are only used during training.

\item \textbf{OGameData250k ---} We also incorporate the OGameData250k dataset, a subset of GameGen-X \cite{Che2024GameGenXIO}, consisting of 226\,k video samples after removing out the test set. Following a setup similar to~\cite{Saito2025SoundReactorFO}, we identified that the original annotations in this dataset primarily describe visual scenes rather than auditory events. Consequently, we discard the original labels for audio conditioning. Instead, we exclusively train using synthetic captions generated via the same pipeline used for VGGSound (Qwen3-Omni on 5-second clips). 
We report our results on the OGameData250k test set utilizing the test split defined in \cite{Saito2025SoundReactorFO}.
\end{itemize}

For evaluation we also use the FoleyBench dataset \cite{Dixit2025FoleyBenchAB}, a high-quality benchmark designed for visually-grounded audio generation. It consists of 5000~curated instances focusing on non-speech and non-music events. Unlike existing large-scale training sets, FoleyBench provides cleaner, more strictly aligned audio-visual pairs, serving as our primary benchmark for assessing generation fidelity and alignment.


\subsection{Training}

First, visual features are extracted from input video sequences using the SynchFormer~\cite{iashin24} encoder at a frame rate of 24\,Hz. Then, these embeddings are projected via a linear layer to match the diffusion transformer's hidden dimension, resulting in a sequence of video condition tokens. 
To accommodate variable video durations, we utilize a validity mask for the real video embeddings. Masked-out positions, representing the unconditioned parts of the sequence, are assigned to a fixed, non-trainable embedding initialized with zero values. 
Additionally, to match the temporal dynamics of the visual data, we compute positional encodings for the video tokens by resampling the RoPE frequencies to align with the video token rate. 
To integrate these tokens, we modify the pre-trained Woosh-Flow multimodal transformer blocks (Section~\ref{sfxfm_t2a}). We introduce a dedicated video modality alongside the existing text and noise latent streams. Specifically, we augment the MultiStream blocks by adding learnable projections for query, key, and value , as well as a distinct FFN specific to the video modality (see above).

Training is initialized from the Woosh-Flow checkpoint. We employ a mixed training strategy, sampling a balanced distribution of approximately 50\% video-audio pairs and 50\% audio-only data. During training, we apply conditioner dropout with a probability of 0.1 randomly masking real video embeddings to improve robustness and facilitate CFG.

\subsection{Distillation}

To accelerate inference for the video-to-audio model, we apply a distillation process highly analogous to the text-to-audio MeanFlow (MF) approach detailed in Section~\ref{sec_t2a_distillation}. We refer to this efficient model as Woosh-DVFlow. By leveraging the teacher model's predictions alongside CFG distillation, we successfully reduce the required NFEs at inference time from 100 to 4 while maintaining a high-fidelity audio generation.

The training objective follows the formulation presented previously, combining the MF distillation loss with the latent adversarial diffusion distillation framework~\cite{sauer24}. However, the DVFlow video-to-audio distillation differs from DFlow in two key aspects. First, training is performed utilizing the VGGSound dataset to better align the student model with training data that presents the three modalities. Second, to mitigate training instabilities introduced by the addition of the video conditioning stream, we empirically found it necessary to increase the adversarial loss weight.

\subsection{Evaluation}
\label{sec:eval_vflow}

To comprehensively assess the performance of our model, we employ a suite of metrics covering audio quality, semantic alignment, and audio-visual synchronization. On top of the FD, KL, and CLAP metrics used in Section~\ref{sec:flow_evaluation}, we also assess:
\begin{itemize}
\item \textbf{Cross-modal alignment ---} We evaluate the alignment between the generated audio and the input conditioning modalities using ImageBind (IB)~\cite{Girdhar23imagebind}, which assesses visual-auditory alignment. We compute the average cosine similarity between the audio embeddings and the embeddings of video frames sampled at 2-second intervals.

\item \textbf{Synchronization analysis ---} We utilize the pre-trained SynchFormer~\cite{iashin24} model to predict the synchronicity between the video and generated audio, following the protocol used in MMAudio~\cite{Cheng2024MMAudioTM}. 
We nonetheless observed significant irregularities with this metric. Specifically, the pre-trained SynchFormer model frequently assigns lower synchronization scores to the ground truth video than to generated samples. Furthermore, we noted that MMAudio achieves exceptionally high scores on this metric; we hypothesize this is because the MMAudio architecture conditions each audio frame directly on interpolated SynchFormer embeddings, effectively optimizing for this specific metric by design. In our experiments on OGameData, while SynchFormer predicts a desynchronization larger than 900\,ms for our model, a manual qualitative inspection of the samples revealed no perceptual sync issues. Consequently, we report this metric only to follow common practice, and with the caveat that it may not reliably reflect perceptual synchronization in this context.
\end{itemize}

We evaluate our method on the test sets of FoleyBench and OGameData250k (see Section~\ref{sec:data_vflow}). Across these datasets, we compare our performance against MMAudio-M~\cite{Cheng2024MMAudioTM}, which serves as a parameter-comparable baseline. Results on the common VGGSound-Test~\cite{Chen20vggsound} are also shown in Appendix~\ref{appendix:results}.


\begin{table}[t]
\centering
\caption{Woosh-VFlow and Woosh-DVFlow results on FoleyBench (top) and OGameData-Test (bottom). We provide the number of latent diffusion model parameters (Params), whether the model is text-conditioned or not (TC), the FD, KL, and CLAP metrics described in Section~\ref{sec:flow_evaluation}, and the task-specific IB and DeSync metrics described in Section~\ref{sec:eval_vflow}.}
\vspace*{2mm}
\label{tab:multi_results}
\setlength{\tabcolsep}{9pt}
\resizebox{\columnwidth}{!}{%
\begin{tabular}{llccrrrrr}
\toprule
\textbf{Dataset} & \textbf{Approach} & \textbf{Params} & \textbf{TC} & \textbf{FD} $\downarrow$ & \textbf{KL} $\downarrow$ & \textbf{IB $\uparrow$} & \textbf{DeSync} $\downarrow$ & \textbf{CLAP $\uparrow$}  \\
\midrule
\multirow[c]{5}{*}{FoleyBench} & Ground truth & - & $\checkmark$ & 0.000 & 0.000 & 0.300 & 0.628 & 0.231 \\
 & MMAudio-M & 621M & $\checkmark$ & 30.480 & 1.983 & \textbf{0.297} & \textbf{0.400} & 0.292 \\
 & Woosh-VFlow & 413M & $\checkmark$ & \textbf{24.136} & \textbf{1.774} & 0.293 & 0.575 & \textbf{0.325} \\
 & Woosh-DVFlow & 413M & $\checkmark$ & 24.422 & 1.834 & 0.278 & 0.647 & 0.303 \\ 
\cmidrule{2-9}
 & MMAudio-M & 621M & $\times$ & 38.517 & 2.466 & 0.268 & \textbf{0.439} & \textbf{0.195} \\
 & Woosh-VFlow & 413M & $\times$ & \textbf{34.382} & \textbf{2.401} & \textbf{0.269} & 0.602 & 0.176 \\
 & Woosh-DVFlow & 413M & $\times$ & \textbf{31.745} & 2.403 & 0.248 & 0.674 & 0.182 \\ 
\hline\hline
\multirow[c]{5}{*}{OGameData} & Ground truth & - & $\checkmark$ & 0.000 & 0.000 & 0.311 & 0.942 & n/a \\
 & MMAudio-M & 621M & $\checkmark$ & 87.186 & 2.267 & 0.215 & 0.927 & \textbf{0.240} \\
 & Woosh-VFlow & 413M & $\checkmark$ & \textbf{11.150} & \textbf{1.006} & \textbf{0.327} & \textbf{0.913} & 0.202 \\
 & Woosh-DVFlow &  413M & $\checkmark$ & \textbf{10.859} & 1.076 & 0.304 & 0.922 & 0.197 \\ 
\cmidrule{2-9}
 & MMAudio-M & 621M & $\times$ & 86.746 & 2.758 & 0.207 & 0.926 & 0.107 \\
 & Woosh-VFlow & 413M & $\times$ & \textbf{11.793} & \textbf{1.376} & \textbf{0.336} & \textbf{0.893} & \textbf{0.123} \\
 & Woosh-DVFlow & 413M & $\times$ & 12.239 & 1.435 & 0.313 & 0.913 & 0.126 \\ 
\bottomrule
\end{tabular}
}
\end{table}

Table~\ref{tab:multi_results} presents the quantitative evaluation of our models compared to the MMAudio-M baseline. 
On the high-quality FoleyBench dataset, our Woosh-VFlow architecture demonstrates strong audio generation capabilities despite having fewer parameters than the baseline (413\,M vs.\ 621\,M). Woosh-VFlow consistently achieves better audio quality, yielding lower FD and KL divergence scores in both text-conditioned and video-only settings. Furthermore, our distilled model, Woosh-DVFlow, successfully maintains highly competitive metrics on FoleyBench while reducing the required inference steps from 100 to 4. We also observe a substantial performance gap in favor of Woosh-VFlow on the OGameData dataset. However, it should be noted that our model was explicitly trained on in-domain gameplay videos (OGameData250k), which is expected to provide a significant advantage in this specific domain compared to the MM-Audio. 

MMAudio achieves the lowest DeSync scores on FoleyBench, notably scoring lower than the actual ground truth videos. As discussed above, we find that this automated metric does not always reliably reflect human-perceived synchronization. Therefore, we encourage readers to visit our supplementary demo page to qualitatively assess the audio-visual synchronization of the generated samples.



\section{Future Work and Applications}
The described foundation models are part of an initial release. They are a solid basis for developing numerous downstream applications in audio generation. For sound editing professionals, we provide a list of potential enhancements and applications below (some of which we already have available internally):

\begin{itemize}
    \item Addition of \textbf{creative controls}, allowing further precision in the generated audio. These may involve conditioning based on precise time-wise editing of acoustic attributes such as loudness or some spectral description. Cross-attention or adaptive layer normalization methods can accomplish this task.

    \item Generation of \textbf{variations} of existing sound inputs. This is useful for, e.g., footsteps in video games. Corruption of sound input latents with additional noise or sampling inversion can be used for this edit.

    \item Enable \textbf{inpainting}, completing a region of audio that temporally stitches with an existing sound. A specific generative model trained to reconstruct only the masked portion of a sound can perform this task.

    \item \textbf{Personalize} a model using one/few samples of audio, allowing to tightly control and modify the audio output while reaching character consistency. Fine-tuning techniques like Dreambooth \cite{ruiz23}, or TokenVerse \cite{garibi25} for further concept disentanglement, have proven to be effective for one-shot and few-shot personalization.
    
    \item \textbf{Morphing} a sound into another using a semantic description of the target sound. To accomplish this, one can for instance sample with a target prompt from noise-corrupted latents computed from some (other) input audio.

    \item \textbf{Loop}-based generation. This can be done by simply looping noise latents prior to sampling.
    
 \end{itemize}

\section{Conclusion}
This report describes Sony AI's Woosh model release 
for high-quality sound effect generation,
mainly addressing text-to-audio and video-to-audio tasks. 
Woosh models trained on public and private datasets were 
benchmarked against other popular open-source models, resulting in
competitive or better performance when evaluated on a public dataset.
When evaluated on private data, Woosh models trained on professional 
sound effect libraries outperformed all other models 
trained on public datasets in terms of Frechet Distance, the only considered model-agnostic metric. This highlights not only the domain shift
present from publicly-available to commercial sound effect 
datasets in terms of audio properties and textual 
annotations, but also the relevance of using the appropriate data
when targeting professional applications.

\printbibliography

\begin{appendices}

\section{Distillation Pseudocode}
\label{appendix:ladd}
Algorithm~\ref{alg_ladd} provides detailed pseudocode for the Woosh-DFLow and Woosh-DVFlow training processes, using the MeanFlow criterion together with latent adversarial diffusion distillation.

\begin{algorithm}[ht]
\caption{Distillation Training with Adversarial Loss}
\label{alg_ladd}
\begin{algorithmic}[1]
\Repeat
    \State \textbf{Sample data and noise:} $x_0 \sim p_{\textrm{data}},\ x_1 \sim \mathcal{N}(0,I)$
    \State \textbf{Train discriminator:}
    \If{$\texttt{training\_step} > 5000$}
        \State \textrm{Unfreeze discriminator}
        \State Sample $(t, r) \sim \texttt{tr\_scheduler}$
        \State Build noisy input: $x_t \gets (1-t)\cdot x_0 + t \cdot x_1$
        \State Student's prediction: $x_r^{\textrm{fake}} \gets x_t - (t-r)\cdot \texttt{student}(x_t, t, r)$
        \State Ground truth: $x_r^{\textrm{true}} \gets (1-r)\cdot x_0 + r \cdot x_1$
        \State Discriminator loss and backpropagation:
        \[
        \begin{aligned}
        &\mathcal{L}_{\textrm{disc}} \gets \textrm{ReLU}\big[1 - \texttt{disc}(x_r^{\textrm{true}})\big] + \text{ReLU}\big[1 + \texttt{disc}(x_r^{\textrm{fake}}.\texttt{detach()})\big] \\
        &\mathcal{L}_{\textrm{disc}}.\texttt{backward()}
        \end{aligned}
        \]
    \EndIf
    \State \textbf{Train generator:}
    \State Freeze discriminator
    \State Sample $(t, r) \sim \texttt{tr\_scheduler}$
    \State Build noisy input: $x_t \gets (1-t)\cdot x_0 + t \cdot x_1$
    \State Get distillation target: $v_{\textrm{tgt}} \gets \texttt{teacher}(x_t, t).\texttt{detach()}$
    \State Student's prediction:
    \[
    u, \tfrac{du}{dt} \gets \texttt{jvp}\bigl(\texttt{student}(x_t, t, r), (v_{\textrm{tgt}}, 1, 0)\bigr)
    \]
    \State Tangent:
    \[
    \begin{aligned}
        &g \gets \bigl(u - v_{\textrm{tgt}} + (t-r)\cdot\tfrac{du}{dt}\bigr).\texttt{detach().clip}(-1,1)
    \end{aligned}
    \]
    \State Mean Flow loss:
    \[
    \mathcal{L}_{\textrm{MF}} \gets \|u - u.\texttt{detach()} + g\|^2
    \]
    \State Adversarial loss: 
    \[
    \begin{aligned}
        &x_r \gets x_t - (t-r)\cdot u \\
        &\mathcal{L}_{\textrm{adv}} \gets -\texttt{disc}(x_r)
    \end{aligned}
    \]
    \State Generator loss and backpropagation:
    \[
    \begin{aligned}
    &\mathcal{L}_{\textrm{gen}} \gets \mathcal{L}_{\textrm{MF}} + 0.5\cdot\mathcal{L}_{\textrm{adv}} \\
    &\mathcal{L}_{\textrm{gen}}.\texttt{backward()}
    \end{aligned}
    \]
\Until{convergence}
\end{algorithmic}
\end{algorithm}

\section{Additional Results}
\label{appendix:results}

In this section, we present supplementary evaluation results for Woosh-VFlow and Woosh-DVFlow on the VGGSound test set. As previously noted in Section~\ref{sec:data_vflow}, evaluating on VGGSound presents inherent challenges due to noisy visual-audio alignments and original dataset labels that are frequently inaccurate or only describe isolated acoustic events. To better understand the models performance, we evaluate generation performance across three distinct conditioning setups: (1) text-conditioned generation using the original dataset labels, (2) text-conditioned generation using our enriched synthetic captions generated via Qwen3-Omni, and (3) an unconditioned (video-only) setup. 

Table~\ref{tab:results:vgg-sound} highlights the significant impact of text conditioning quality on the final audio generation. Utilizing the synthetic Qwen3-Omni captions provides a much more accurate reflection of the complex auditory scenes, helping to mitigate the limitations of the dataset's original noisy labels. By comparing these scenarios, we demonstrate how higher-quality text conditioning guides the model toward outputs that are more semantically and temporally aligned with the video.

\begin{table}[t]
\centering
\caption{Woosh-VFlow and Woosh-DVFlow results on VGG-Sound.}
\vspace*{2mm}
\label{tab:results:vgg-sound}
\begin{tabular}{lccrrrrr}
\toprule
Approach & Params & TC & FD & KL & IB & DeSync & CLAP \\
\midrule
Ground truth & - & $\checkmark$ & -0.000 & 0.000 & 0.331 & 0.630 & 0.198 \\
MMAudio-M & 621M & $\checkmark$ & 30.539 & \textbf{1.413} & \textbf{0.329} & \textbf{0.449} & 0.224 \\
Woosh-VFlow & 413M & $\checkmark$ & \textbf{25.711} & 1.691 & 0.282 & 0.633 & \textbf{0.278} \\
Woosh-DVFlow & 413M & $\checkmark$ & 25.469 & 1.722 & 0.278 & 0.673 & 0.235 \\ 
\midrule
MMAudio-M & 621M & Recap & 31.445 & 1.540 & \textbf{0.309} & \textbf{0.448} & \textbf{0.166} \\
Woosh-VFlow & 413M & Recap & \textbf{19.442} & \textbf{1.338} & 0.270 & 0.620 & 0.161 \\
Woosh-DVFlow & 413M & Recap & 22.343 & 1.472 & 0.259 & 0.675 & 0.150 \\ 
\midrule
MMAudio-M & 621M & $\times$ & 30.528 & \textbf{1.846} & \textbf{0.320} & \textbf{0.436} & \textbf{0.162} \\
Woosh-VFlow & 413M & $\times$ & \textbf{20.837} & 2.118 & 0.286 & 0.577 & 0.136 \\
Woosh-DVFlow & 413M & $\times$ & 24.336 & 2.192 & 0.266 & 0.655 & 0.128 \\ 
\bottomrule
\end{tabular}
\end{table}

\end{appendices}

\end{document}